\documentclass{article}
\usepackage{spconf,amsmath,graphicx,hyperref, multirow, booktabs, graphicx, stfloats}
\usepackage{rotating}
\usepackage{tabularx, amssymb, wrapfig}

\title{Hierarchical Activity Recognition and Captioning from Long-Form Audio}

\name{
  \begin{minipage}{\linewidth}
    \centering
    Peng Zhang, Qingyu Luo, Philip J.B. Jackson, Wenwu Wang
  \end{minipage}
}

\address{Centre for Vision, Speech and Signal Processing (CVSSP), University of Surrey, U.K.}

\begin{document}

\maketitle

\begin{abstract}
Complex activities in real-world audio unfold over extended durations and exhibit hierarchical structure, yet most prior work focuses on short clips and isolated events. To bridge this gap, we introduce \textit{MultiAct}, a new dataset and benchmark for multi-level structured understanding of human activities from long-form audio. MultiAct comprises long-duration kitchen recordings annotated at three semantic levels (activities, sub-activities and events) and paired with fine-grained captions and high-level summaries. We further propose a unified hierarchical model that jointly performs classification, detection, sequence prediction and multi-resolution captioning. Experiments on MultiAct establish strong baselines and reveal key challenges in modelling hierarchical and compositional structure of long-form audio. A promising direction for future work is the exploration of methods better suited to capturing the complex, long-range relationships in long-form audio.

\end{abstract}


\vspace{-10pt}
\section{Introduction}
\vspace{-10pt}
\label{sec:intro}

While audio understanding has seen increasing attention in recent years, the recognition of complex human activities purely from auditory signals remains largely understudied~\cite{shin2025comprehensive, wijngaard2025audio, huh2025epic}. Despite recent progress in sound event detection and captioning, most existing datasets~\cite{gemmeke2017audio, chen2020vggsound, drossos2020clotho} and models~\cite{hou2023cooperative, mei2024wavcaps, hu2025pseldnets} are constrained to short clips (e.g., 10–30 seconds), which typically capture only isolated acoustic events or brief action snippets. As a result, the understanding of long-form, structured human activities remains largely unexplored. However, in real-world settings, meaningful activities often span extended durations and comprise sequences of temporally related events~\cite{damen2020epic}. Effective modelling of this long-form acoustic structure, particularly procedural relationships among events, is essential for deeper semantic understanding but is not adequately addressed in current research~\cite{kong2020panns, hou2023audio}.

To advance long-form activity understanding, it is essential to model the structural complexity of human behaviour while addressing the limitations of existing audio datasets~\cite{mei2022automated}. Drawing inspiration from the ecological psychology framework~\cite{roger1955children, luo2021moma}, human activity can be analysed across three levels: behavior settings, behavior episodes, and atomic physical units. A behavior setting denotes a stable activity pattern bounded by specific spatial, temporal, and social constraints. In a kitchen scenario, a high-level activity such as {\small\texttt{Cleanup}} functions as a behavior setting that structures goal-oriented behavior episodes, including sub-activities such as {\small\texttt{scrubbing a pan}}. These episodes are further composed of fine-grained atomic physical units, manifested in the auditory domain as primitive sound events such as {\small\texttt{water running}} or {\small\texttt{metal collision}}. Modelling such a hierarchy is therefore crucial for robust semantic representation and reasoning over complex auditory sequences. However, widely used audio datasets, such as AudioSet~\cite{gemmeke2017audio}, VGGSound~\cite{chen2020vggsound} and Clotho~\cite{drossos2020clotho}, are not well suited for this purpose. As shown in Table~\ref{tab:dataset}, these datasets consist of short clips with flat, event-level annotations, lacking the temporal scope, sequential relationships, and hierarchical organisation required to study compositional semantics and high-level activity.

\begin{table}[t]
\centering
\footnotesize
\caption{Comparison of audio datasets by annotation level.}
\setlength{\tabcolsep}{2pt}
\begin{tabular}{l c c c c c}
\toprule
Dataset & Scale & Avg. Dur & Event & Sub-activity & Activity \\
\midrule
AudioSet~\cite{gemmeke2017audio} & $\sim$5800h & 10s & \checkmark & $\times$ & $\times$ \\
VGGSound~\cite{chen2020vggsound}  & $\sim$550h & 10s & \checkmark & $\times$ & $\times$ \\
Clotho~\cite{drossos2020clotho}   & $\sim$24h & 15s--30s & (Captions) & $\times$ & $\times$ \\
EPIC-SOUNDS~\cite{huh2025epic} & $\sim$100h & $\sim$4.9s & \checkmark & $\times$ & $\times$ \\
\midrule
\textbf{MultiAct} & $\sim$9h & -- & \checkmark & \checkmark & \checkmark \\
\bottomrule
\end{tabular}
\label{tab:dataset}
\end{table}

To address this challenge, we introduce the \textit{MultiAct} dataset, a new benchmark for multi-level structured understanding of human activities from long-form audio. MultiAct is partially derived from EPIC-SOUNDS~\cite{huh2025epic}, where we re-annotate a curated subset and reorganise it into long, contiguous segments with hierarchical temporal labels and paired textual descriptions. The MultiAct dataset has three distinguishing features: (1) hierarchical annotations at three semantic levels: high-level activities, sub-activities and events, enabling the study of compositional structure and sequential reasoning; (2) long-form audio recordings of up to 50 minutes, allowing models to capture extended temporal dependencies and activity flow; (3) dual natural-language descriptions for each segment, comprising fine-grained captions and high-level summaries, supporting multi-resolution language generation. To the best of our knowledge, MultiAct is the first audio dataset with these features.

Furthermore, we introduce a unified hierarchical framework for understanding long-form audio activities. We explore tasks such as hierarchical activity recognition, sub-activity sequence prediction and captioning at multiple resolutions. Experiments on the MultiAct dataset demonstrate strong performance across tasks and reveal key challenges in modelling structured long-form audio. All resources are available at: {\footnotesize\href{https://github.com/PennyZhang9/MultiAct}{\texttt{github.com/PennyZhang9/MultiAct}}}.

\begin{figure}[!t]
  \centering
  \includegraphics[width=\columnwidth]{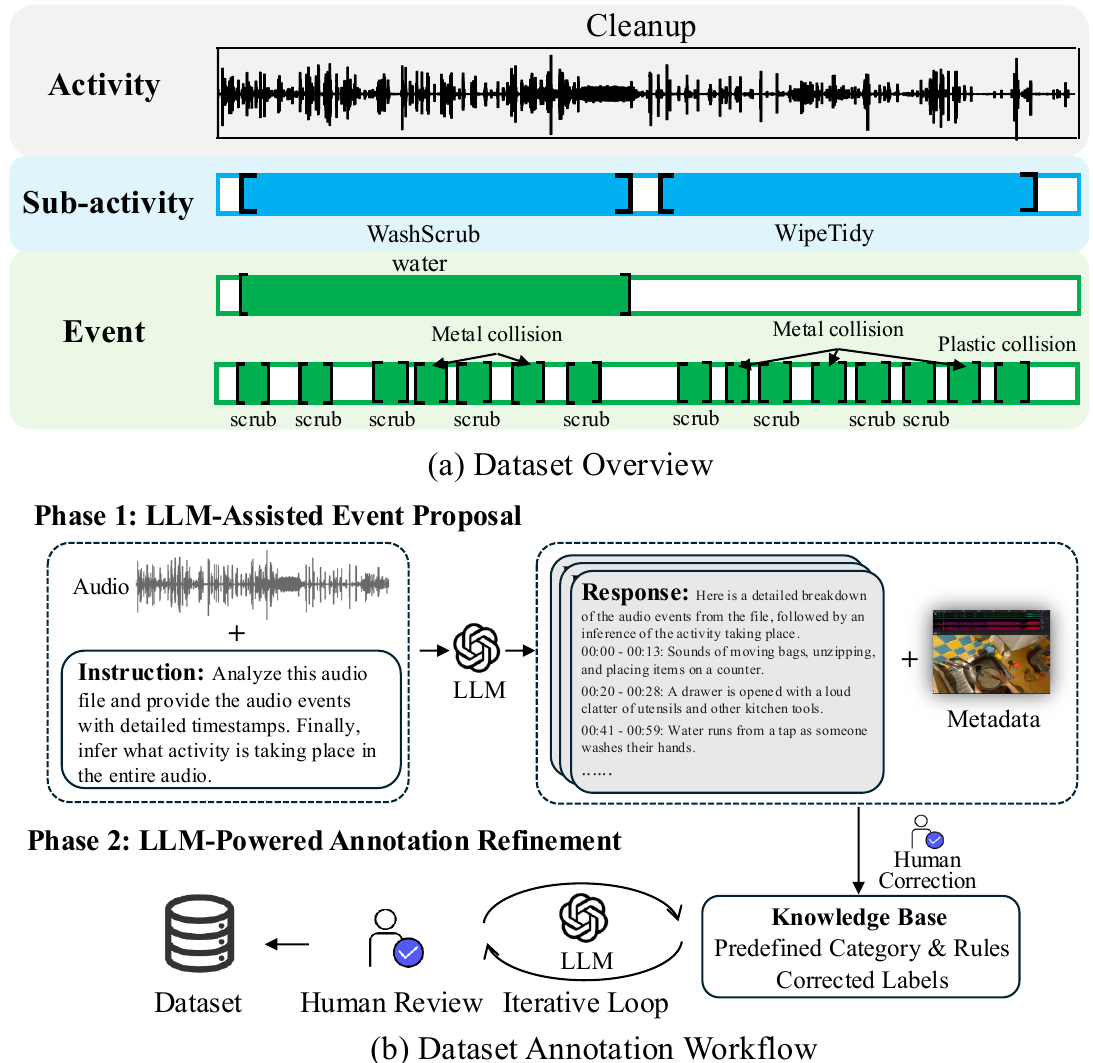}
  \vspace{-10pt}
  \caption{The hierarchical structure of \textit{MultiAct} and LLM-assisted annotation pipeline.}
  \label{fig:overview}
\end{figure}

\vspace{-12pt}
\section{Dataset}
\vspace{-10pt}
\label{sec:dataset}

The MultiAct dataset is derived from recordings of individuals engaged in cooking activities within a kitchen environment. This dataset is structured in a three-level hierarchy grounded in activity partonomy, providing rich temporal annotations at the levels of activity, sub-activity, and event. As shown in Fig~\ref{fig:overview}(a), each high-level activity (e.g., {\small\texttt{Cleanup}}) is composed of multiple sub-activities (e.g., {\small\texttt{WashScrub}}, {\small\texttt{WipeTidy}}), which in turn consist of a sequence of fine-grained sound events (e.g., {\small\texttt{scrub}}, {\small\texttt{metal collision}}, {\small\texttt{water}}). This hierarchical structure enables the modelling of procedural semantics across different temporal resolutions.

MultiAct comprises 8.97 hours of long-form audio from 17 kitchen environments sampled from EPIC-SOUNDS~\cite{huh2025epic}, augmented with additional hierarchical annotations. It includes 51 activity instances across 3 coarse activity classes, with durations ranging from 15.7s to 3164s (mean 628s). At the sub-activity level, it contains 472 annotated instances spanning 12 categories, with durations from 0.57s to 572s (mean 63.6s). For event-level modelling, the taxonomy reuses and refines 44 classes, comprising 7312 instances with durations between 0.2s and 100s (mean 4.91s). In addition to temporal annotations, MultiAct also includes fine-grained captions aligned with short-term acoustic events and high-level summaries describing the overall activity, supporting text generation tasks at multiple granularities. Our MultiAct annotations are released under the Creative Commons Attribution (CC BY) license. The accompanying audio is sourced from EPIC-SOUNDS~\cite{huh2025epic}, a derivative of EPIC-KITCHENS~\cite{damen2020epic}, and remains subject to the original non-commercial EPIC-KITCHENS Research License.

Annotations are constructed through a two-phase Large Language Model (LLM)-assisted workflow, as shown in Fig.~\ref{fig:overview}(b). In Phase 1, we prompt GPT-4o~\cite{hurst2024gpt}, with long-form audio content and task-specific instructions to generate a structured annotation draft. Although the output may include hallucinations or boundary errors due to the audio complexity and long duration, it serves as a rich initialisation for downstream refinement. In Phase 2, these drafts are iteratively corrected by human annotators, guided by metadata and a knowledge base of predefined categories and labelling rules. This human-in-the-loop process is particularly critical given the dataset’s multi-level annotation scheme and ensures high-quality, consistent labels across all semantic levels.

\vspace{-14pt}
\section{Method}
\vspace{-10pt}
\label{sec:method}

\begin{figure*}[!t]
  \centering
  \includegraphics[width=0.96\linewidth, height=0.37\linewidth]{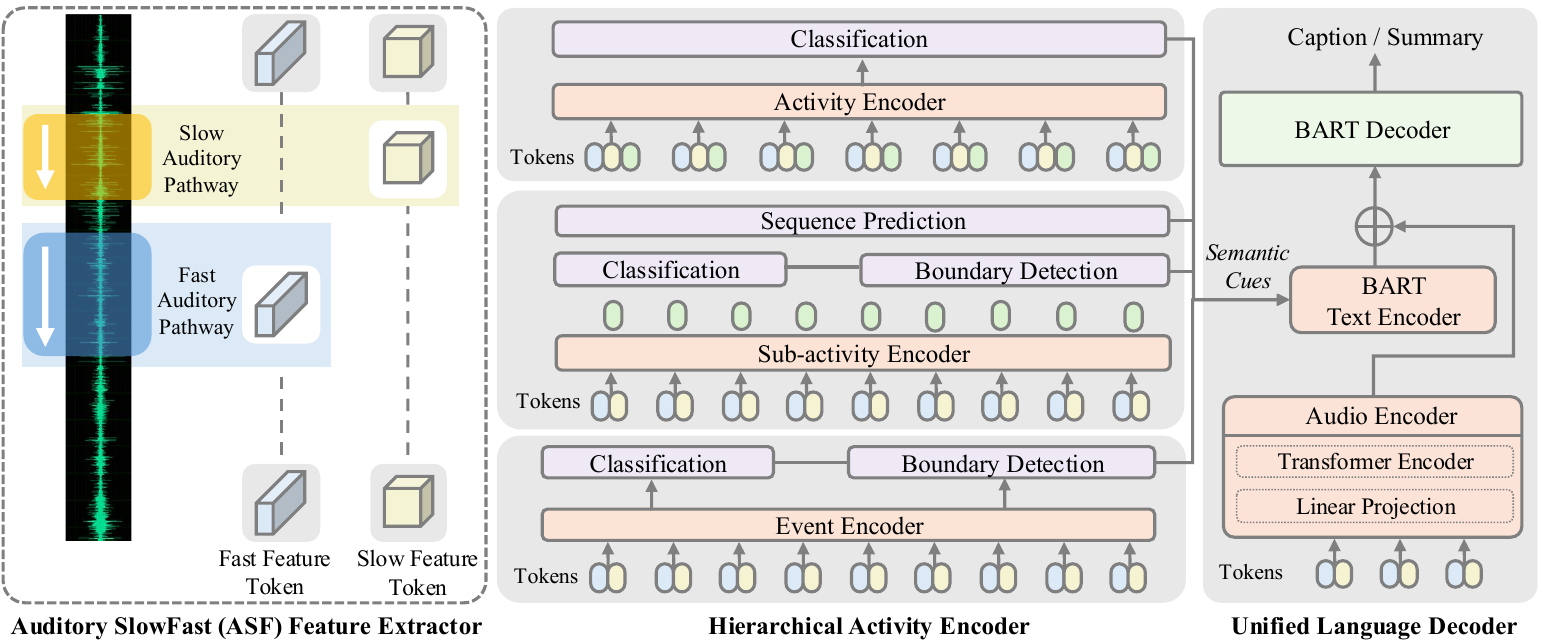}
  \vspace{-6pt}
  \caption{Overview of the proposed architecture.}
  \label{fig:waveform}
\end{figure*}

As illustrated in Fig.~\ref{fig:waveform}, we propose a unified hierarchical framework for long-form audio activity recognition and captioning, consisting of three components: Auditory SlowFast (ASF)~\cite{kazakos2021slow} feature extractor, multi-level hierarchical activity encoder and unified language decoder for caption and summary generation. 

A SlowFast-style encoder with synchronised fast and slow paths extracts temporally aligned acoustic features: the fast path captures transient cues at high resolution; the slow path aggregates longer-range spectral structure. The fused outputs form tokenised, frame-level representations for all downstream tasks.

To model semantic structure at multiple temporal resolutions, we design a hierarchical encoder composed of three task-specific components. The \textbf{event encoder} performs fine-grained event classification and boundary detection. The \textbf{sub-activity encoder} models mid-level temporal structure via sub-activity classification, boundary detection, and sequence prediction, and fuses its embeddings with frame-level ASF features for activity-level prediction. The \textbf{activity encoder} predicts high-level procedural categories. All three encoders operate on a shared audio token representation extracted from the ASF backbone and are trained separately to produce level-specific outputs.

To generate natural-language outputs, we adopt a unified BART-based captioning decoder~\cite{lewis2019bart}. Audio features from the ASF encoder are first linearly projected and processed by a multi-layer Transformer audio encoder. Optionally, semantic cues from any level of the hierarchical encoder (e.g., sub-activity sequences or activity labels) are encoded by the pretrained BART text encoder to provide textual conditioning. The audio and text encoder outputs are then concatenated and fed to the BART decoder, which autoregressively generates either fine-grained captions or high-level summaries, depending on the task.

\vspace{-12pt}
\section{Experiments}
\vspace{-10pt}
\label{sec:exp}

In this section, we evaluate our approach on four tasks using the MultiAct dataset: (i) \textbf{Hierarchical Activity Classification}, predicts one label at each level (activity, sub-activity, event), given a temporally bounded audio segment; (ii) \textbf{Audio Activity Detection}, produces class-labelled temporal segments on untrimmed audio; (iii) \textbf{Audio Activity Sequence Prediction}, predicts the ordered sub-activity sequence to capture temporal dynamics and intent from a segment; and (iv) \textbf{Audio Activity Captioning}, generates either a fine-grained, time-ordered caption or a high-level summary for the long-form audio. In all experiments, the ASF backbone~\cite{kazakos2021slow} pretrained on EPIC-SOUNDS~\cite{huh2025epic} serves as a frozen feature extractor, producing 2304-dimensional frame-level audio features that are fed to the task-specific models.

\vspace{-12pt}
\subsection{Hierarchical Activity Classification}
\vspace{-8pt}
\label{ssec: classification}

The purpose of the activity classification task is to evaluate whether a model offers a comprehensive understanding of activities from auditory information across multiple semantic levels. Given a temporally bounded audio segment (with start and end times), the model predicts one label at each of the activity, sub-activity and event levels. This is formulated as three parallel single-label, multi-class classification tasks.

\noindent\textbf{Baselines.} We implement three baselines to study architectural choices. Event-level classification is performed by training a linear layer on top of the ASF audio features. For sub-activity and activity classification, we freeze ASF and stack two Bidirectional Gated Recurrent Unit (BiGRU) layers (256 hidden units per direction) on top to encode temporal dynamics, followed by task-specific attention. ASF-Atten applies self-attention pooling~\cite{zhu2018self} to the BiGRU outputs for sub-activity classification. ASF-CrossAtten introduces a cross-attention module~\cite{vaswani2017attention} and jointly trains sub-activity and activity heads to capture hierarchical dependencies. We report five standard metrics: Top-1 and Top-5 accuracy, mean Per-Class Accuracy (mPCA), mean Average Precision (mAP), and mean Area Under the Curve (mAUC). Higher values indicate better performance for all metrics. For Activity (3 classes), Top-5 is undefined and therefore omitted (cells marked N/A).

\begin{table}[t]
  \centering
  \footnotesize
  \caption{Results of the classification baseline models (\%, $\uparrow$).}
  \vspace{0.1em} 
  \scalebox{0.8}{ 
    \begin{tabular}{cccccccc}
    \toprule
    Split & Level & Model & Top-1 & Top-5 & mPCA  & mAP & mAUC \\
    \midrule
    \multirow{5}[10]{*}{Val} & Event & ASF   & 69.0  & 93.6  & 30.1  & 41.6  & 93.5 \\
\cmidrule{2-8}          & \multirow{2}[4]{*}{Sub-activity} & ASF-Atten & 50.0  & \textbf{92.6}  & 55.2  & \textbf{64.9}  & 85.5 \\
\cmidrule{3-8}          &       & ASF-CrossAtten & \textbf{51.9}  & 88.9  & \textbf{49.7}  & 65.9  & \textbf{87.5} \\
\cmidrule{2-8}          & \multirow{2}[4]{*}{Activity} & ASF-Atten & 60.0  & N/A   & 60.7  & \textbf{79.4}  & 87.2 \\
\cmidrule{3-8}          &       & ASF-CrossAtten & \textbf{66.7}  & N/A   & \textbf{61.9}  & 72.7  & 84.6 \\
    \midrule
    \multirow{5}[10]{*}{Eval} & Event & ASF   & 67.2  & 92.4  & 33.1  & 41.5  & 91.9 \\
\cmidrule{2-8}          & \multirow{2}[4]{*}{Sub-activity} & ASF-Atten & \textbf{51.3}  & \textbf{83.3}  & \textbf{47.4}  & \textbf{41.9}  & \textbf{77.5} \\
\cmidrule{3-8}          &       & ASF-CrossAtten & 42.3  & 74.4  & 33.8  & 37.0  & 76.4 \\
\cmidrule{2-8}          & \multirow{2}[4]{*}{Activity} & ASF-Atten & 83.3  & N/A   & 83.3  & \textbf{94.4}  & \textbf{95.8} \\
\cmidrule{3-8}          &       & ASF-CrossAtten & \textbf{83.3}  & N/A   & \textbf{83.3}  & 72.2  & 70.8 \\
    \bottomrule
    \end{tabular}%
    }
  \label{tab:table_2}%
\end{table}%

\noindent\textbf{Results.} Table \ref{tab:table_2} shows the performance across all three semantic levels on both validation and evaluation splits. ASF-CrossAtten performs best on activity classification (83.3\% Eval), while ASF-Atten excels at Sub-activity, benefiting from temporal modelling.

\vspace{-12pt}
\subsection{Audio Activity Detection}
\vspace{-8pt}
\label{ssec: detection}

In this experiment, we further evaluate the temporal detection task, which aims to detect activity segments along the temporal axis and assign each segment a class label (e.g., sub-activity or event). This is formulated as a multi-class temporal detection problem that jointly predicts segment boundaries (start and end) and the class of each detected segment.

\noindent\textbf{Baselines.} We adapt ActionFormer~\cite{zhang2022actionformer}, a strong video action detection model, to the audio domain. During training, we extract ASF features using 2-s windows with a 200-ms stride, and feed them to ActionFormer for temporal proposal generation and classification. Performance is reported as mAP across multiple Intersection-over-Union (IoU) thresholds.

\noindent\textbf{Results.} As shown in Table~\ref{tab:table_3}, sub-activity detection consistently outperforms event detection, with higher AP across both splits. However, performance drops notably at higher IoU thresholds. For instance, on the evaluation set, the sub-activity AP falls from 41.8\% at an IoU of 0.1 to 22.0\% at an IoU of 0.5, suggesting challenges in precise boundary localization.

\begin{table}[t]
  \centering
  \caption{Results of the detection baseline models (\%, $\uparrow$).}
  \vspace{0.1em} 
  \scalebox{0.8}{ 
    \begin{tabular}{cccccccc}
    \toprule
    \multirow{2}[4]{*}{Split} & \multirow{2}[4]{*}{Level} & \multicolumn{6}{c}{AP at Different IoU Thresholds} \\
\cmidrule{3-8}          &       & @0.1  & @0.2  & @0.3  & @0.4  & @0.5  & Mean \\
    \midrule
    \multirow{2}[4]{*}{Val} & Event & 17.0  & 14.7  & 12.8  & 11.2  & 9.8  & 13.1  \\
\cmidrule{2-8}          & Sub-activity & 44.3  & 41.0  & 30.8 & 25.3 & 24.3 & 33.1 \\
    \midrule
    \multirow{2}[4]{*}{Eval} & Event & 16.5 & 15.7 & 15.0 & 13.6 & 12.5 & 14.6 \\
\cmidrule{2-8}          & Sub-activity & 41.8 & 37.3 & 32.2 & 26.0 & 22.0 & 31.9 \\
    \bottomrule
    \end{tabular}%
    }
  \label{tab:table_3}%
\end{table}%

\begin{table}[t]
  \centering
  \caption{Results of the sequence prediction models for sub-activities (\%, $\downarrow$).}
  \vspace{0.1em} 
  \scalebox{0.8}{ 
    \begin{tabular}{ccccccc}
    \toprule
    \multicolumn{1}{c}{\multirow{2}[4]{*}{Split}} & \multicolumn{6}{c}{AER on Different Training Context Length} \\
    \cmidrule{2-7}          
     & 2 & 3 & 4 & 6 & 8 & \multicolumn{1}{c}{Full} \\
    \midrule
    Val & 66.7 & 72.2 & 75.9 & 81.5 & 88.9 & 79.6 \\
    \midrule
    Eval & 69.2 & 74.4 & 75.6 & 87.2 & 87.2 & 80.8 \\
    \bottomrule
    \end{tabular}%
    }
  \label{tab:table_4}%
\end{table}%

\vspace{-12pt}
\subsection{Audio Activity Sequence Prediction}
\vspace{-8pt}
\label{ssec: sequence prediction}

In audio datasets, activity sequence prediction has been rarely defined or systematically studied in prior work. Leveraging MultiAct’s dense, time-aligned annotations and rich sub-activity taxonomy, we formalise the task as sequence prediction: given a bounded audio segment, predict the ordered sub-activity sequence. The model infers both which sub-activities appear and their order without explicit segment boundaries, emphasising procedural structure and long-range dependencies rather than local boundary detection.

\noindent\textbf{Baselines.} We use ASF audio features as input to a Conformer-based encoder~\cite{gulati2020conformer}. The features are first projected with a linear layer. Adopting the convolution kernel size of 31 from the original study, our encoder is configured with 8 layers and processes these features to produce rich contextual representations. A final linear layer maps these representations to label logits, and the model is trained with the Connectionist Temporal Classification (CTC) objective~\cite{graves2006connectionist} to predict sub-activity sequences. During inference, we apply sliding window decoding to handle long-form audio. We evaluate the Activity Error Rate (AER) as a function of training context length, varying the context from 2, 3, 4, 6, to 8 sub-activities, and also including the entire sequence (“Full”).

\noindent\textbf{Results.} Table~\ref{tab:table_4} reveals that shorter contexts (2–4) produce the lowest AER, while error increases with longer contexts (6–8) and only partially improves with Full. These results highlight the importance of local context and the difficulty of modelling long-range dependencies, underscoring sequence prediction as a challenging open problem.

\begin{table}[t]
  \centering
  \footnotesize
  \caption{Results of the fine-grained captioning (C) and high-level summarization (S) baseline models (\%, $\uparrow$).}
  \vspace{0.1em} 
  \scalebox{0.8}{ 
    \begin{tabular}{cccccccc}
    \toprule
    \multicolumn{1}{c}{Split} & \multicolumn{1}{c}{Task} & \multicolumn{1}{c}{Method} & \multicolumn{1}{c}{BLEU1} & \multicolumn{1}{c}{BLEU4} & \multicolumn{1}{c}{METEOR} & \multicolumn{1}{c}{ROUGE-L} & \multicolumn{1}{c}{CIDEr} \\
    \midrule
    \multicolumn{1}{c}{\multirow{4}[8]{*}{Val}} & \multirow{2}[4]{*}{C} & Rule  & 24.0 & 4.9  & 11.4 & 20.0 & 3.8 \\
\cmidrule{3-8}          &       & Hierarchical & \textbf{39.2} & \textbf{10.8} & \textbf{15.8} & \textbf{28.5} & \textbf{16.2} \\
\cmidrule{2-8}          & \multirow{2}[4]{*}{S} & Rule  & 23.7  & 7.99  & 11.7 & 26.9 & 9.3 \\
\cmidrule{3-8}          &       & Hierarchical & \textbf{28.4} & \textbf{9.8}  & \textbf{13.5}  & \textbf{32.5}  & \textbf{24.1} \\
    \midrule
    \multicolumn{1}{c}{\multirow{4}[8]{*}{Eval}} & \multirow{2}[4]{*}{C} & Rule  & 16.8 & 2.7  & \textbf{13.0} & 20.2 & 2.8 \\
\cmidrule{3-8}          &       & Hierarchical & \textbf{17.8} & \textbf{3.4}  & 9.6  & \textbf{23.0} & \textbf{20.1} \\
\cmidrule{2-8}          & \multirow{2}[4]{*}{S} & Rule  & 17.3 & \textbf{7.5}  & 8.8  & 20.7 & 2.2 \\
\cmidrule{3-8}          &       & Hierarchical & \textbf{21.8} & 7.3  & \textbf{11.0} & \textbf{28.3} & \textbf{11.1} \\
    \bottomrule
    \end{tabular}%
    }
  \label{tab:table_5}%
\end{table}%

\vspace{-12pt}
\subsection{Audio Activity Captioning}
\vspace{-8pt}
\label{ssec: captioning}

In this experiment, we investigate whether explicitly modelling hierarchical relations among events, sub-activities and activities improves the coherence and semantic fidelity of captions for long-form audio. This task focuses on generating natural language descriptions of long-form audio segments at two levels: fine-grained captioning and high-level summarisation.

\noindent\textbf{Baselines.} To establish a captioning baseline, we adopt a dual-encoder, BART-decoder~\cite{lewis2019bart} architecture. ASF-derived audio features are linearly projected to the BART hidden size and encoded by a Transformer audio encoder (10 layers, 8 attention heads). In parallel, the BART text encoder processes structured cues from multi-level activity encoders. We then concatenate the audio and text encoder outputs and pass them to the BART decoder to generate captions. Training uses teacher forcing with token-level cross-entropy; inference uses beam search (beam size 4) with a length penalty and trigram blocking. We evaluate the performance with BLEU-1/4~\cite{papineni2002bleu}, METEOR~\cite{banerjee2005meteor}, ROUGE-L~\cite{lin2004rouge}, and CIDEr~\cite{vedantam2015cider}.

\noindent\textbf{Results.} Table~\ref{tab:table_5} reports the results for fine-grained captioning (C) and high-level summarisation (S) on the validation and evaluation splits. The hierarchical model outperforms the rule-based baseline on most metrics, with the largest gains in CIDEr and ROUGE-L. Summarisation generally achieves higher BLEU, METEOR, and ROUGE than captioning, whereas captioning attains the highest CIDEr on the evaluation split. Overall, performance is higher on summarisation than on fine-grained captioning, underscoring the challenge of generating temporally precise and semantically rich descriptions for long-form audio.

\vspace{-12pt}
\section{Conclusion}
\vspace{-10pt}
\label{sec:conclusion}

We present MultiAct, a first-of-its-kind benchmark for structured long-form audio, offering hierarchical annotations (activities, sub-activities, events) together with paired fine-grained captions and high-level summaries. Alongside the dataset, we introduce a unified hierarchical model for joint recognition and captioning. Experiments on MultiAct validate the approach and highlight persistent challenges in modelling semantic and procedural structure in long-form audio. We hope this work catalyses sustained progress in high-level activity understanding from long-form audio.

\vspace{-12pt}
\section{Acknowledgment}
\vspace{-10pt}
\label{sec:acknowledgment}

We thank Pablo Martínez-Nuevo, Sven Ewan Shepstone and Jon Francombe (Bang \& Olufsen A/S) for valuable discussions. This research was supported by Bang \& Olufsen A/S.

\small      
\begingroup
\bibliographystyle{IEEEtran} 
\bibliography{refs}

\end{document}